\documentclass[english,prl,twocolumn,showpacs,superscriptaddress]{revtex4-1}
\usepackage[T1]{fontenc}
\usepackage[latin9]{inputenc}
\usepackage{color}
\usepackage{amssymb}
\usepackage{graphicx}
\usepackage{esint}

\makeatletter
\@ifundefined{textcolor}{}
{%
 \definecolor{BLACK}{gray}{0}
 \definecolor{WHITE}{gray}{1}
 \definecolor{RED}{rgb}{1,0,0}
 \definecolor{GREEN}{rgb}{0,1,0}
 \definecolor{BLUE}{rgb}{0,0,1}
 \definecolor{CYAN}{cmyk}{1,0,0,0}
 \definecolor{MAGENTA}{cmyk}{0,1,0,0}
 \definecolor{YELLOW}{cmyk}{0,0,1,0}
 }

\usepackage[bookmarks]{hyperref}

\makeatother

\usepackage{babel}
\begin{document}

\title{Localized phase structures growing out of quantum fluctuations in
a quench of tunnel-coupled atomic condensates}

\author{Clemens Neuenhahn }

\email{clemens.neuenhahn@physik.uni-erlangen.de}

\affiliation{Friedrich-Alexander-Universität Erlangen-Nürnberg, Institute for
Theoretical Physics II, Staudtstr. 7, 91058 Erlangen, Germany}

\author{Anatoli Polkovnikov}

\affiliation{Department of Physics, Boston University, 590 Commonwealth Avenue,
Boston, Massachusetts 02215, USA}

\author{Florian Marquardt}

\affiliation{Friedrich-Alexander-Universität Erlangen-Nürnberg, Institute for
Theoretical Physics II, Staudtstr. 7, 91058 Erlangen, Germany}
\begin{abstract}
We investigate the relative phase between two weakly interacting 1D
condensates of bosonic atoms after suddenly switching on the tunnel-coupling.
The following phase dynamics is governed by the quantum sine-Gordon
equation. In the semiclassical limit of weak interactions, we observe
the parametric amplification of quantum fluctuations leading to the
formation of breathers with a finite lifetime. The typical lifetime
and density of the these 'quasibreathers' are derived employing exact
solutions of the classical sine-Gordon equation. Both depend on the
initial relative phase between the condensates, which is considered
as a tunable parameter.
\end{abstract}
\maketitle
Dynamical instabilities can amplify spatial field fluctuations drastically.
If the instability provides sufficient energy, even quantum zero-point
fluctuations can trigger the formation of macroscopic field patterns.
For instance, in some cosmological scenarios of the inflationary stage
of the universe, following a slow-roll, a scalar inflaton field performs
oscillations around the minimum of the corresponding inflaton potential.
Thereby, resonant spatial fluctuations are amplified parametrically.
For a rather generic class of potentials these can end in long-lived,
local concentrations of energy, so-called 'oscillons' (e.g.,\cite{2003_Gleiser,2011_AminInflation,2010_Amin3D,2008_Farhi}).

In this work, we argue that analogous non-equilibrium phenomena should
be observable in experiments with a pair of weakly interacting quasi-1D
clouds of cold, bosonic atoms \cite{2007_Hofferbeth,2011_Betz}. After
suddenly turning on the tunnel coupling between the condensates, the
dynamics of the relative phase field $\hat{\phi}$ is governed by
the integrable quantum sine-Gordon model \cite{2007_AnatoliSG} 
\begin{eqnarray}
\frac{d^{2}\hat{\phi}}{dt^{2}}-\frac{d^{2}\hat{\phi}}{dx^{2}}+\frac{m^{2}}{\beta}\sin\beta\hat{\phi}=0,\label{EquationsOfMotion}
\end{eqnarray}
where the 'mass' $m$ depends on the tunnel amplitude, such that $m(t)=m\Theta(t)$
for this quench. The phase has been rescaled, and for weak interactions
$\beta\ll1$.

The sine-Gordon model (SGM) is one of the most prominent prototypical
models of low-dimensional condensed-matter systems. Currently, quenches
in the SGM are under intense investigation (e.g., \cite{2007_GritsevPRL,2008_Grandi,2010_MitraSG,2010_CazalillaNJP}).
For instance, only recently the amplification of density inhomogeneities
after a quench to the special Luther-Emmery point ($\beta^{2}=4\pi$),
was demonstrated \cite{2010_Foster}. 

\begin{figure}
\includegraphics[width=1\columnwidth]{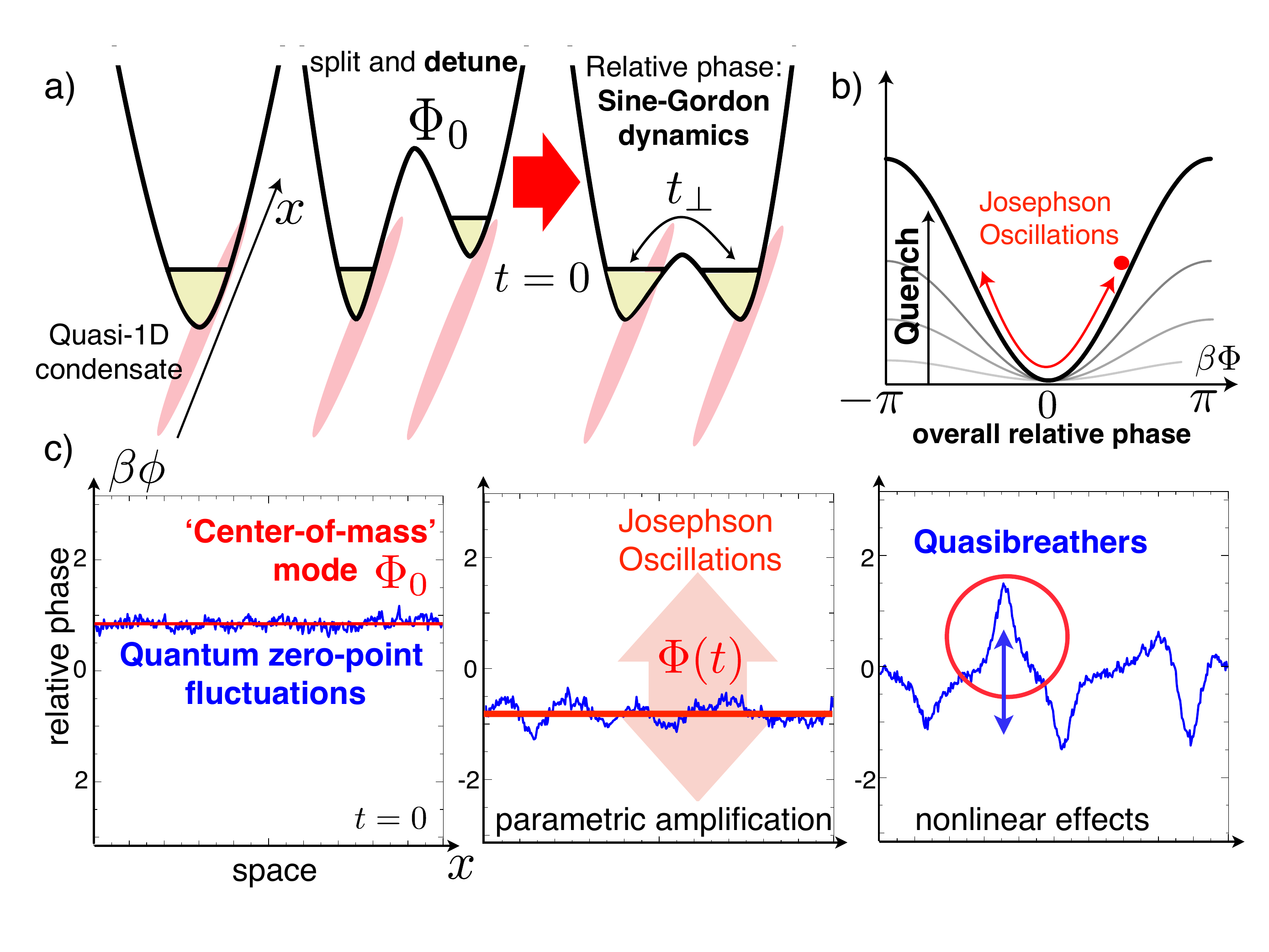}\caption{a) Proposed experimental protocol. A single quasi-1D condensate is
split and the global phase is tuned to $\Phi_{0}$. Switching on the
tunnel-coupling, the dynamics of the relative phase obeys the quantum
sine-Gordon equation. b) At short times, the global phase performs
Josephson oscillations c) Single run of TWA (see main text) with $\beta\Phi_{0}=0.25\pi$.
Spatial quantum fluctuations are amplified parametrically. Eventually,
the non-linearity of the sine-Gordon equation kicks in and 'quasibreathers'
(breathers with a finite lifetime) form.\label{Figure1}}
\end{figure}

Here, we propose that the spatially averaged value $\Phi$ of the
relative phase field is tuned to some value $\Phi_{0}$ right before
the quench (e.g., this might be achieved by slightly tilting the transversal
double-well potential confining the BECs, cf. Fig.$\,$\ref{Figure1}a).
The subsequent phase dynamics can be directly observed in matter wave
interference experiments \cite{2007_Hofferbeth,2011_Betz}. 

At short times, $\Phi$ will perform Josephson oscillations (Fig.$\,$\ref{Figure1}b)
according to $d^{2}\Phi/dt^{2}=-\beta^{-1}m^{2}\sin\beta\Phi$. However,
these are linearly unstable \cite{1999_Greene,2005_Bochoule} in the
presence of inhomogeneous quantum fluctuations which are parametrically
amplified at certain wave lengths (Fig.$\,$\ref{Figure1}c). As we
will show in this paper, due to this modulation instability at some
later point the dynamics becomes fully nonlinear and one observes
the formation of sharply localized and oscillating patterns in the
phase-field (Fig.$\,$\ref{Figure1}c). It will be shown below that
they can be related to particular exact solutions of the classical
SGM obtained from the Bäcklund transformation (following \cite{1978_Scott}).
These 'quasibreathers' (QBs) have a finite lifetime (in contrast to
the well-known breather solutions), and we find good evidence that
a quasi-equilibrium steady-state with a finite density of such excitations
develops at long times\textcolor{red}{.} 

For the considered quench, due to the intrinsic instability, a simple
approximation of the SGM in terms of non-interacting, massive phonons
is invalid, even though $\beta\ll1$. Instead, we employ the truncated
Wigner-approximation (e.g., \cite{2010_AnatoliTWA}) (TWA). The basic
idea behind TWA is to simulate classical field equations, but with
quantum-mechanical fluctuations as stochastic initial conditions.
In the limit of $\beta\rightarrow0$, TWA is expected to become reliable
as it can be shown that $\beta$ plays the role of an effective Planck's
constant \cite{2010_AnatoliTWA}. It describes correctly the linear
dynamics during the parametric amplification. Once QBs form, occupation
numbers are already large such that a semiclassical description (provided
by TWA) should continue to remain valid. The great advantage of TWA
is that it serves snapshots (Fig.$\,$\ref{Figure1}c) of the phase
field $\phi(x,t)$ run-by-run, which can be compared directly to the
experimental observations. 

In the following, we introduce the model and demonstrate that the
proposed setup should be well within the reach of present experiments.
After a discussion of the numerical findings obtained from TWA, we
will introduce the analytical 'quasibreather' solutions of the classical
SGM. We demonstrate that these solutions are well suited to explain
the main physical features as predicted by TWA. Eventually, we argue
that a statistical analysis of $\phi(x,t)$, experimentally obtained
at a single time $t$ per run, could reveal the distinctive signature
of QBs. 

\emph{Model}. \textendash{} It was shown \cite{2007_AnatoliSG} that
on scales larger than the condensate healing length $\xi_{h}$, the
dynamics of the relative phase between condensates $\hat{\phi}\equiv(\hat{\phi}_{1}-\hat{\phi}_{2})/\sqrt{2}$
(and of the density field $\hat{\Pi}$, fulfilling $[\hat{\Pi}(x),\hat{\phi}(x')]=i\hbar\delta(x-x')$)
is governed by the quantum SGM. In the following, we use the notation
in \cite{2007_AnatoliSG}. After rescaling the fields ($\hat{\phi}\mapsto\sqrt{\pi/K}\hat{\phi}$
and $\hat{\Pi}\mapsto\sqrt{K/\pi}\hat{\Pi}$), the Hamiltonian reads
\begin{eqnarray}
\hat{H}=\frac{\hbar v_{s}}{2}\int dx\,\left[\hat{\Pi}^{2}+(\partial_{x}\hat{\phi})^{2}\right]-\frac{m^{2}}{\beta^{2}}\int dx\,\cos\beta\hat{\phi}.\label{eqSG}
\end{eqnarray}
The repulsive short-range interaction, characterized by the Luttinger
parameter $K$, enters $\beta=\sqrt{2\pi/K}$. For repulsive bosons
$K\in[1,\infty]$. In the limit of small interactions\cite{2007_AnatoliSG}
($K\rightarrow\infty$), $K$ is connected to the Lieb-Liniger parameter
$\gamma=m_{B}g/\hbar^{2}\rho_{0}$ via $K=\pi/\sqrt{\gamma}$ (where
$g$ is the interaction strength, $m_{B}$ the mass of the bosons
and $\rho_{0}$ the mean density). Furthermore, there are the sound
velocity $v_{s}=v_{F}/K$ (with $v_{F}=\hbar\pi\rho_{0}/m_{B}$),
and the effective mass $m=\sqrt{2t_{\perp}\rho_{0}}\beta$, where
the tunnel-amplitude $t_{\perp}$ between the condensates enters.
The relevant length scale in the sine-Gordon model, determining the
width of breathers and solitons, is set by $\sqrt{\hbar v_{s}}/m$.
The sine-Gordon model should be a sound description of the system
as long as $\sqrt{v_{s}\hbar}/m\gg\xi_{h}\sim1/\rho_{0}\sqrt{\gamma}$.
Finally, the Josephson frequency is given by $\omega_{J}=m\sqrt{v_{s}/\hbar}$
\cite{2003_Bochoule}, determining the period of the Josephson oscillations
for $\beta\Phi_{0}\lesssim1$. In the following, we set $v_{s}=\hbar=1$
. 

\emph{Experimental Realizability}. \textendash{} In the limit of weak
interactions ($\beta\ll1$), the stiffness of the condensate becomes
small. To avoid the breakdown of the hydrodynamical description, the
mean condensate density $\rho_{0}$ must be large enough. In fact,
the condition $\sqrt{v_{s}\hbar}/m\gg\xi_{h}$ translates into a lower
bound for $\beta$ (and therefore into a lower bound for the interaction
strength): $\beta^{2}\gg4\rho_{0}^{-1}\sqrt{t_{\perp}m_{{\rm B}}}/\hbar$.
Within the present experimental setups the condensate density can
be widely tuned. E.g., in \cite{2010_KruegerPRL} the 1D density of
${\rm {Rb}^{87}}$ atoms (with $m_{B}=1.44\cdot10^{-25}{\rm {kg}}$)
ranges from $\rho_{0}=3-100{\mu{\rm {m}}}^{-1}$. For a transverse
trap frequency $\omega_{\perp}=2\pi\times4{\rm {kHz}}$, with $g=2\hbar\omega_{\perp}a_{s}$
and $a_{s}=5.31\cdot10^{-9}{\rm {m}}$, we have $K=9-52$ and therefore
small values of the sine-Gordon parameter $\beta=\mathcal{O}(10^{-1})$
are well achievable. Eventually, the tunnel-amplitude can be tuned
between $t_{\perp}/\hbar=[\mathcal{O}(10^{2})-\mathcal{O}(10^{4})]/{\rm {s}}$
\cite{2007_Hofferbeth}. Thus, for large densities $\rho_{0}=100\mu{\rm {m}^{-1}}$
the lower bound on $\beta$ is $\mathcal{O}(10^{-1})$ as well. In
all numerical simulations, we use $\beta=0.1$. We conclude that after
some fine-tuning of the experimental parameters, the proposed setup
should be within reach.

\begin{figure}
\includegraphics[width=1\columnwidth]{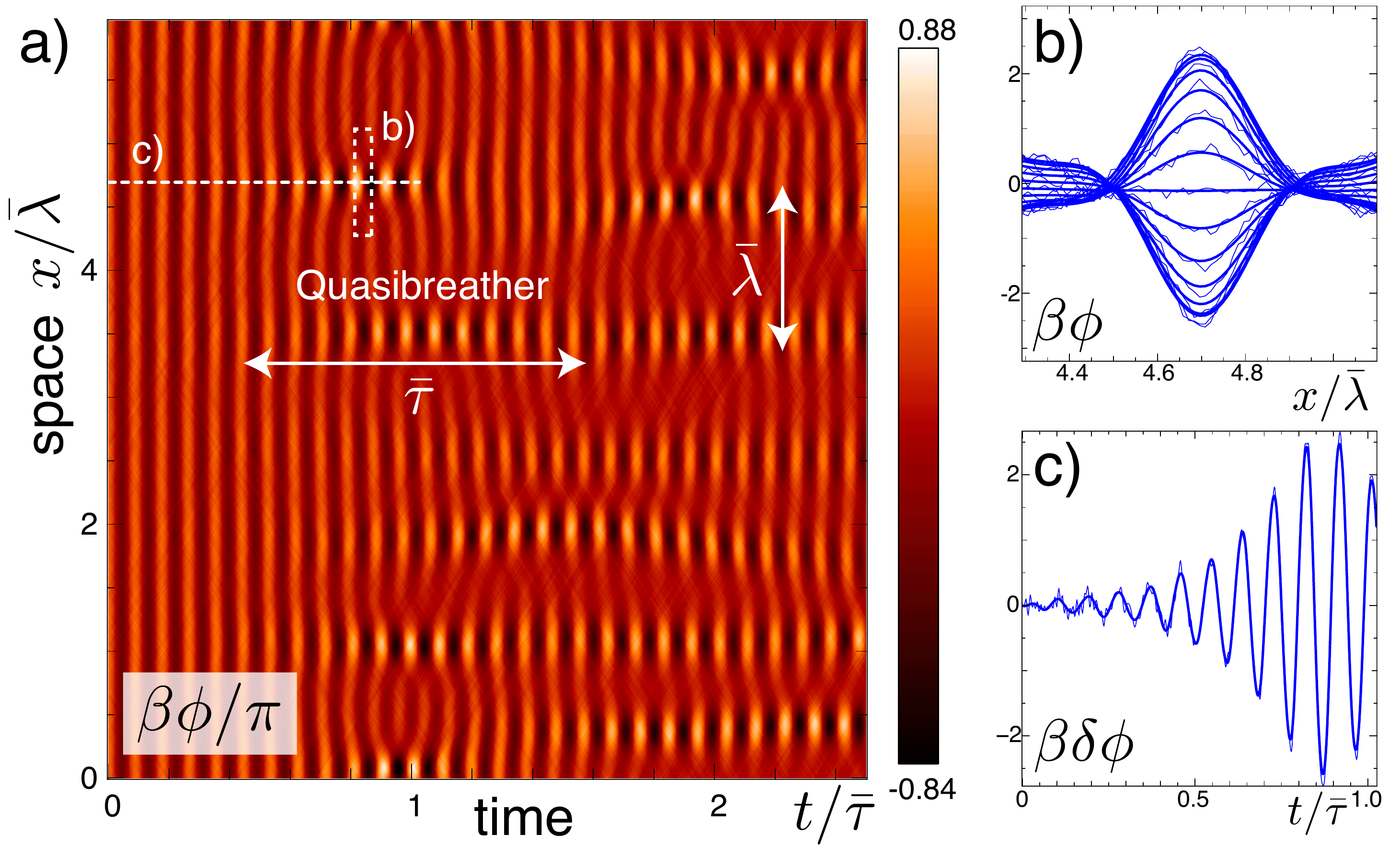}\caption{a) Space-time plot of the phase-field for a single simulation run.
b) Spatial cut through the quasibreather highlighted in a) over one
breather period (thin blue line; thick blue line: filtering out irrelevant
short wavelength fluctuations). c) Temporal cut through the emerging
QB taken at its center. $\delta\phi=\phi-\Phi$ is obtained by subtracting
the spatially averaged field $\Phi=\frac{1}{L}\int_{0}^{L}dx\,\phi$.
One observes the exponential amplification of initial quantum fluctuations.
Here, $\beta\Phi_{0}=0.3\pi$.\label{Figure2}}
\end{figure}
\emph{Numerical results} \textendash{} Before the quench, the phase
field $\phi$ is given by the offset $\Phi_{0}$ plus the small zero-point
fluctuations of the free theory {[}Eq.$\,$(\ref{eqSG}) with $m=0${]}.
Within TWA, the fluctuations of the $\phi$ and $\Pi$ modes are initialized
according to their Gaussian Wigner-distribution \cite{2010_AnatoliTWA},
omitting the $q=0$ mode. 

This initial state should be experimentally achievable. One should
start from the ground state at strong tunnel coupling ($m\gg1$),
where the relative phase $\phi(x)=0$. During a slow linear rampdown
of the tunnel-amplitude, all modes $\phi_{q\neq0}$ will follow their
time-dependent groundstate. It can be shown that the global phase
$\Phi$ at the end of this process obeys $\beta^{2}\langle\Phi^{2}\rangle\sim1/\xi_{h}\rho_{0}$,
which remains small as long as the number of bosons within the healing
length is large. An additional potential tilt will produce a fixed
phase offset $\Phi\simeq\Phi_{0}$. We will focus on rather small
$\beta\Phi_{0}\lesssim1$. For $\beta\Phi_{0}\rightarrow\pi$, in
addition to breathers one observes the formation of solitons. Finally,
we note that this 'phase-tuning' becomes impossible for infinite systems.
This is due to the logarithmic divergence of phase fluctuations with
system size $L$ in the massless groundstate, i.e., $\beta^{2}\langle\delta\hat{\phi}^{2}\rangle\sim\beta^{2}\ln L$.
However, these are largely suppressed in the semiclassical limit $\beta\ll$1.
We always choose the system size such that the initial overall relative
phase is well defined. 

\begin{figure}
\includegraphics[width=1\columnwidth]{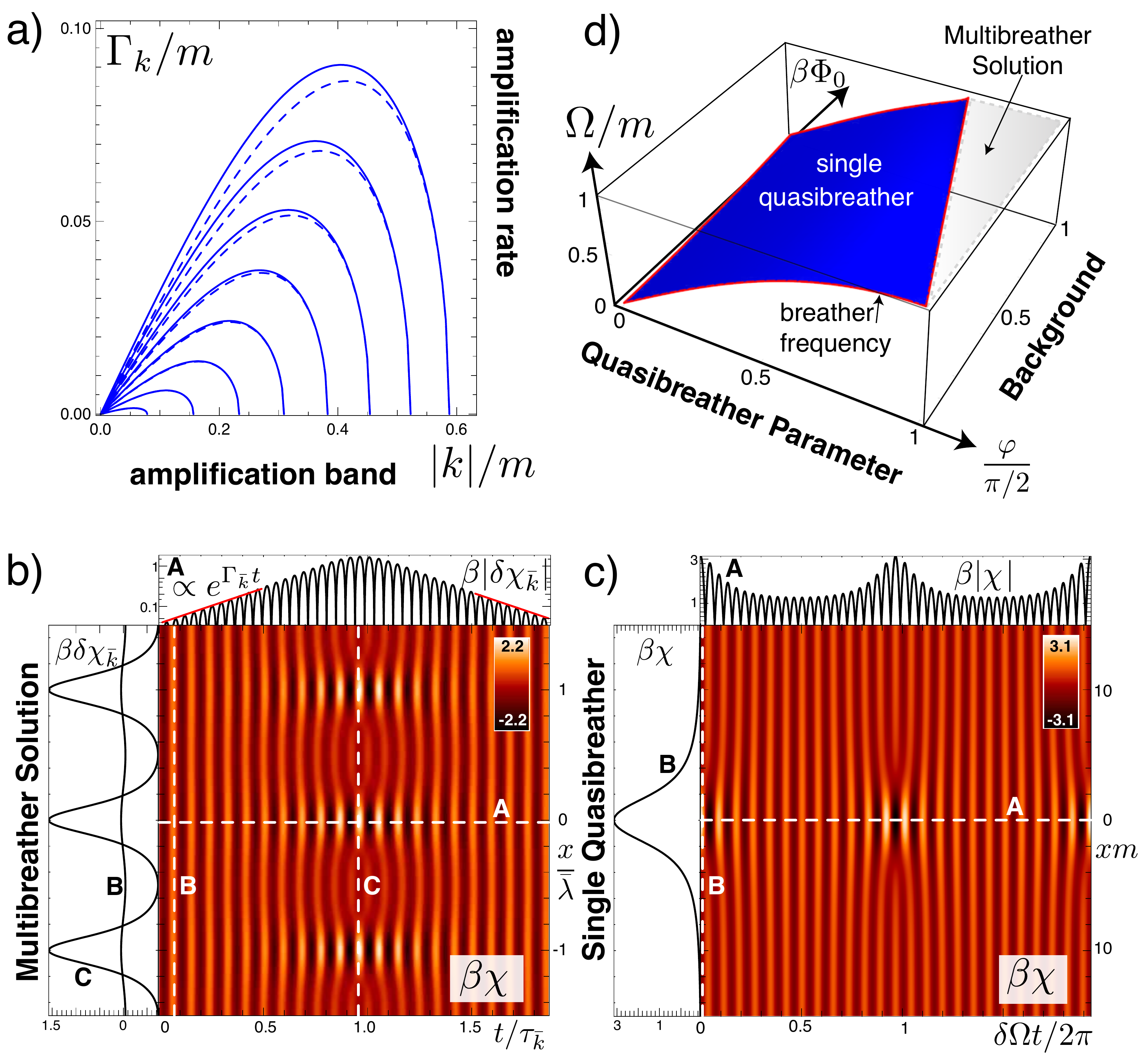}\caption{Phononic zero-point fluctuations are amplified parametrically driven
by the overall phase $\Phi$ performing Josephson oscillations. a)
Corresponding amplification rate $\Gamma_{k}$ of phonons with wavenumber
$k$ for $\beta\Phi_{0}/\pi\in0.05-0.4$ according to Eq.$\,$(\ref{AmplificationRate})
(dashed line). Solid lines show the rate obtained from the 'multibreather'
solution Eq.$\,$(\ref{eq:2-1}). b) Plot of a (spatially periodic)
'multibreather' solution reducing to a driven phonon with wavenumber
$k=m\beta\Phi_{0}/\sqrt{8}$ for $t/\tau_{\bar{k}}\rightarrow-\infty$.
Cut A shows the parametric amplification of this phonon (B) driven
by the oscillating background (we plot $\delta\chi_{\bar{k}}$, substracting
this background). They provide 'seeds' for the formation of Quasibreathers
(C) with a lifetime $\tau_{\bar{k}}$. Here, $\beta\Phi_{0}=0.3\pi$.
c) Plot of a 'single quasibreather' (periodic in time) with $\beta\Phi_{0}=0.3\pi$
and $\varphi=1.02$. d) Mean frequency $\Omega\equiv m-\delta\Omega$
of a 'single quasibreather' ($|\sin\varphi|<|\cos\frac{\beta\Phi_{0}}{2}|$)
depending on the amplitude of the background oscillations and the
parameter $\varphi$. For $\Phi_{0}=0$, one restores the unperturbed
breather frequency $\mbox{\ensuremath{\Omega}=}m\sin\varphi$.\label{Figure3}}
\end{figure}

In Fig.$\,$\ref{Figure2}, the phase field for a single run of TWA
is shown. At small times, the field is dominated by $\Phi(t)$ performing
ordinary Josephson oscillations. At this point, the inhomogeneous
part $\delta\phi(x,t)$ of $\phi=\Phi+\delta\phi$ can be treated
as a small perturbation. The modes obey $\partial_{t}^{2}\phi_{k}+(k^{2}+m^{2}\cos[\beta\Phi(t)])\phi_{k}\simeq0$
for all $k\neq0$, i.e., these are phonons with a periodically modulated
mass. Modes with $|k|\in[0,m\sin|\beta\Phi_{0}/2|]$ are parametrically
amplified yielding $\phi_{k}(t)\sim e^{\Gamma_{k}t}$. The amplification
rates $\Gamma_{k}$ for this linear regime (here displayed for $\beta\Phi_{0}\lesssim1$)
(see Fig.$\,$\ref{Figure3}a) 
\begin{eqnarray}
2\Gamma_{k} & \simeq & k\sqrt{\sin^{2}\left(\beta\Phi_{0}/2\right)-k^{2}/m^{2}},\label{AmplificationRate}
\end{eqnarray}
were found in \cite{1999_Greene}, neglecting the damping of the driving
$\Phi$-mode. After some time, nonlinear interactions between the
amplified modes become important and lead to the formation of sharply
localized oscillating structures (here denoted as 'quasibreathers'),
which constitute the main phenomenon discussed in our paper. Once
these localized oscillations get out of phase with respect to the
background oscillations of $\Phi$, their energy is depleted again.
One arrives at a steady-state, where QBs are randomly created and
decay, with a typical lifetime $\bar{\tau}$ and a mean spatial distance
$\bar{\lambda}$ (cf. Fig.$\,$\ref{Figure2}). All these statistical
quantities crucially depend on the initial value $\Phi_{0}$. It turns
out that these localized modulations in the stochastic phase field
can be connected to certain exact solutions of the SGM. These solutions,
to be discussed in the following, are standing breathers riding on
top of a homogeneous and oscillating background. We will demonstrate
that this set of solutions is well suited to describe the numerical
observations and provides analytical insight into the dependence of
$\bar{\lambda}$ and $\bar{\tau}$ on $\Phi_{0}$. 

\emph{Quasibreathers}. \textendash{}
 A Bäcklund transformation (see for instance \cite{1978_Scott}) allows
to 'add' (anti)solitons to a given solution of the SGM, here taken
to be the spatially homogeneous solution $\Phi(t)$ with $\Phi(0)=\Phi_{0}$
and $\dot{\Phi}(0)=0$. Adapting the approach of~\cite{1978_Scott},
we obtain quasibreather-solutions by adding a (standing) soliton and
the corresponding antisoliton to $\Phi$. A standing SGM-breather
is characterized by its amplitude (e.g., \cite{1979_Maki2}). The
solutions here depend in addition on the amplitude of the underlying
background oscillations $\Phi_{0}$ (furthermore, there is a minor
dependence on the precise value of the relative phase $\Delta_{0}$
between breather and background at $t=0$). They have the form 
\begin{eqnarray}
\chi(x,t) & = & \frac{4}{\beta}\arctan\left[\mathcal{G}(x,t;\Phi_{0},\Delta_{0},\varphi)\right]+\Phi(t).\label{eq0}
\end{eqnarray}
For $\Phi_{0}=0$, the parameter $\varphi\in[0,\pi/2]$ determines
the unperturbed breather frequency $m\sin\varphi$ and ${\rm max}\chi=\beta^{-1}(2\pi-4\varphi)$.
An explicit expression for $\mathcal{G}$ and a detailed discussion
of Eq.$\,$(\ref{eq0}) are given in the supplement. Here, we focus
on the most relevant features in the limit $\beta\Phi_{0}\lesssim1$.

The central observation is that by placing a breather on top of the
oscillating background, its amplitude becomes time-dependent (see
Figs.$\,$\ref{Figure3}b,c). The background $\Phi(t)$ amplifies
the breather while the frequency of the latter decreases. This is
due to the fact that the effective curvature of the cosine-potential
decreases for larger field amplitudes. Eventually, both run out of
phase and, subsequently, the breather gets damped. The relative phase-drift
occurs at a frequency 
\begin{eqnarray}
\delta\Omega(\varphi,\Phi_{0}) & = & \frac{m\sqrt{\left|\cos^{2}\left(\beta\Phi_{0}/2\right)-\sin^{2}\varphi\right|}\cos\varphi}{1+\sin\varphi}.\label{eq:2-1}
\end{eqnarray}
As long as $\sin\varphi<\cos(\beta\Phi_{0}/2)$, $\chi$ describes
a single breather whose amplitude is modulated with a period $2\pi/\delta\Omega$
(Fig.$\,$\ref{Figure3}c). This constitutes a stable QB solution
of the SGM. As the background oscillates at frequency $m$, one can
understand $\Omega\equiv m-\delta\Omega$ as the mean frequency of
the quasibreather (Fig.$\,$\ref{Figure3}d).

For $\sin\varphi\rightarrow\cos(\beta\Phi_{0}/2)$, however, the period
diverges. In fact, it turns out that for $\sin\varphi>\cos(\beta\Phi_{0}/2)$,
$\chi$ is \emph{periodic in} \emph{space} rather than in time (Fig.$\,$\ref{Figure3}b).
It describes a set of quasibreathers at distance $2\pi/k=2\pi m^{-1}\left|\cos^{2}\left(\beta\Phi_{0}/2\right)-\sin^{2}\varphi\right|^{-1/2}$. 

These 'multibreather' solutions have remarkable properties. For $t\rightarrow-\infty$,
they reduce to a phonon $\delta\chi_{k}$ with wavenumber $k$, superimposed
on the oscillating background. It is amplified parametrically, yielding
\begin{eqnarray}
\delta\chi_{k} & \propto & e^{\Gamma_{k}t}{\rm \zeta}(t)\cos kx,\label{Eq3}
\end{eqnarray}
where $\zeta(t)$ is a periodic function and the amplification rate $\Gamma_{k}\equiv\delta\Omega(\varphi(k),\Phi_{0})$.
This rate is in agreement with the results in Eq.$\,$(\ref{AmplificationRate})
for small $\beta\Phi_{0}$. At later times, the characteristic breather
peaks form (Fig.$\,$\ref{Figure3}c), exist during a time set by
$\tau_{k}=4/\Gamma_{k}$ and decay again for $t\rightarrow\infty$.
Note that $\chi$ therefore describes the prototypical formation of
QBs out of fluctuations in a single mode $\phi_{k}$. 

Our numerical analysis shows (see below) that one can infer the properties
of typical QBs observed within the stochastic TWA from these 'ideal'
solutions $\chi$. In a given run, zero-point fluctuations in \emph{all}
modes are present. However, in the weakly interacting limit ($\beta\ll1$),
the parametric instability automatically filters out modes with $k\approx\bar{k}$,
where $\bar{k}\simeq m\beta\Phi_{0}/\sqrt{8}$ denotes the maximally
amplified mode. Thus, the typical \emph{distance} between QBs is roughly
given by $\bar{\lambda}=2\pi/\bar{k}$. Although the strictly periodic
'multibreather' solution $\chi$ is not directly observed, individual
QBs are well described by $\chi$ as long as their amplitude is large
enough compared to the noisy background. Typical QBs decay after a
\emph{lifetime} $\bar{\tau}\equiv\tau_{\bar{k}}$ (cf. Fig.$\,$\ref{Figure3}c).
When their amplitude is depleted down to the noise level, the exact
solution ceases to be relevant. Then the background oscillations initiate
the amplification process again, leading to the formation of new QBs.
This explains the observed stochastic creation and annihilation of
QBs at large times $\Gamma_{\bar{k}}t\gg1$.

\emph{Statistical Analysis}. \textendash{} 
\begin{figure}
\includegraphics[width=1\columnwidth]{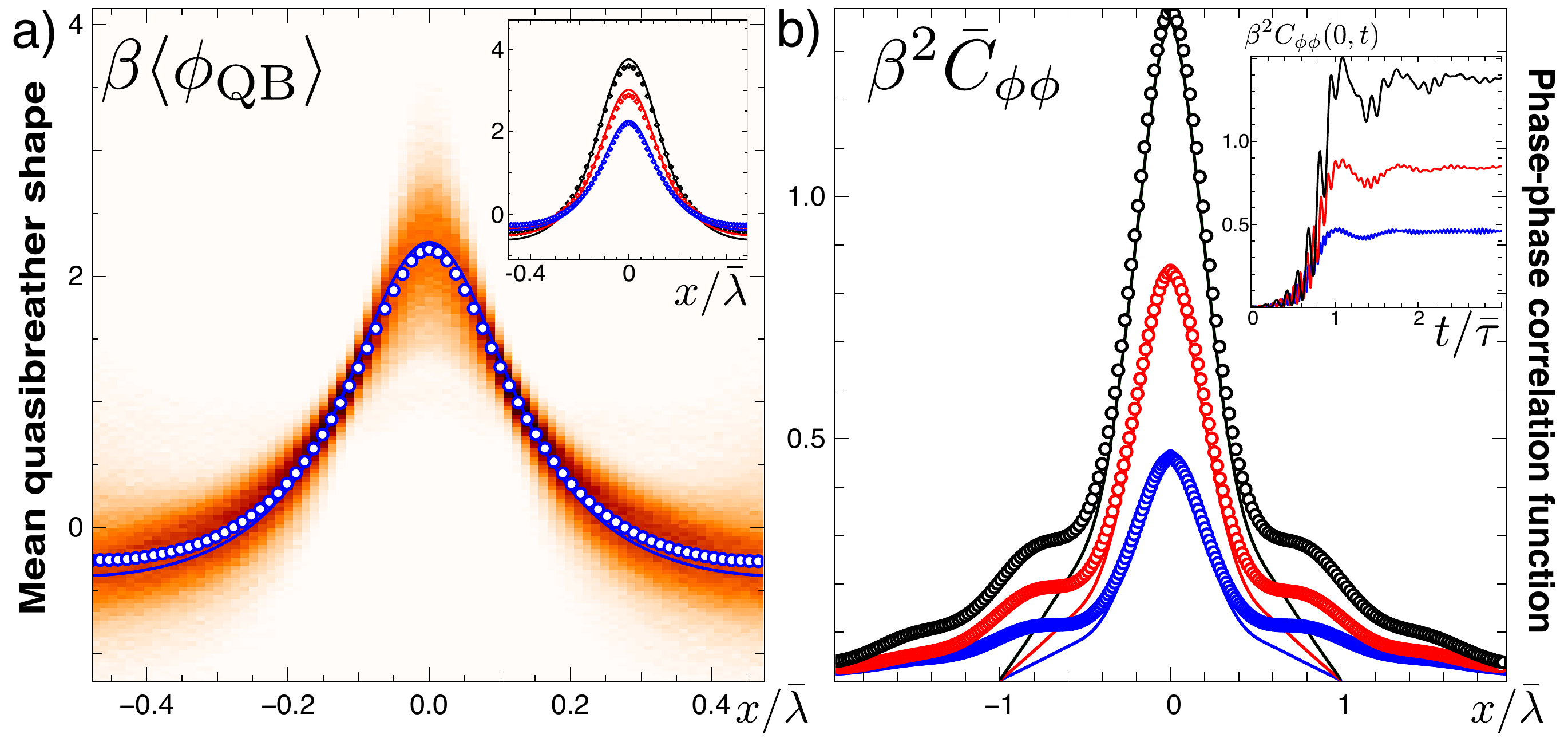}\caption{a) Mean shape of QBs taken at their maximal value, obtained from TWA
(blue dots) and full distribution of shapes (color plot). We track
QBs numerically run-by-run in the interval $t/\bar{\tau}\in[0.5,4]$
for $\beta\Phi_{0}/\pi=0.3$. Comparison to the analytical solution
$\chi(x,0)$ with $\bar{k}$ (blue, solid line; $\Delta_{0}=0$) shows
almost perfect agreement. Inset) Various $\beta\Phi_{0}/\pi=0.3,0.4,0.5$
(from bottom to top). No fit-parameter enters. b) Correlation function
$C_{\phi\phi}$, averaged over time $t/\bar{\tau}\in[2.25,3]$ (dots)
with $\beta\Phi_{0}/\pi=0.3,0.4,0.5$ (from bottom to top). Locally,
it agrees well with Eq.$\,$(\ref{eq:}). Inset shows $C_{\phi\phi}(0,t)$,
demonstrating that the field enters a steady state for $t/\bar{\tau}\gtrsim1$. }

\label{Figure4}
\end{figure}
These predictions agree well with the statistical analysis of the
quench within TWA. For every run, we numerically track all QBs, showing
their shape $\phi_{{\rm QB}}(x)$ at the time of maximum amplitude
(Fig.$\,$\ref{Figure4}a). A comparison of the mean QB-shape $\langle\phi_{{\rm QB}}\rangle$
and $\chi$ (with $k=\bar{k}$) shows excellent agreement. Note that
no fit parameter enters here. 

Finally, Fig.$\,$\ref{Figure4}b shows the equal-time correlation
function $C_{\phi\phi}(x,t)=\langle\hat{\phi}(x,t)\hat{\phi}(0,t)\rangle-\langle\hat{\phi}(0,t)\rangle^{2}$
evaluated with TWA. For $t/\bar{\tau}>1$, it saturates, indicating
that the field enters a statistical steady state. One can find a decent
approximation for $C_{\phi\phi}$ at large times $t\gg\bar{\tau}$,
assuming that $\phi(x,t)$ can be represented as a sum of independent
QBs, at an average density $1/\bar{\lambda}\bar{\tau}$ in the $(x,t)$-plane:
\begin{eqnarray}
C_{\phi\phi} & \approx & \int_{-\bar{\lambda}/2}^{\bar{\lambda}/2}\frac{dx_{0}}{\bar{\lambda}}\int_{-\bar{\tau}/2}^{\bar{\tau}/2}\frac{dt_{0}}{\bar{\tau}}\,\tilde{\chi}(x;x_{0},t_{0})\tilde{\chi}(0;x_{0},t_{0}),\label{eq:}
\end{eqnarray}
which describes the core part of $C_{\phi\phi}$ fairly well (cf.
Fig.$\,$\ref{Figure4}). Here, a single QB from the 'multibreather'
solution centered at $x_{0}$ enters: $\tilde{\chi}(x;x_{0},t_{0})=\Theta(\frac{\bar{\lambda}}{2}-|x-x_{0}|)\chi(x-x_{0},t_{0};\bar{k})$
($\Theta$ denotes the heavy-side step function). 

This correlation function is directly accessible in experiments and
should distinctively reveal the presence of QBs. Moreover, one could
simply perform a direct statistical analysis of $\phi$ (cf. Fig.$\,$\ref{Figure4}a). 

\emph{Summary}. \textendash{} We predict the formation of localized
modulations in the relative phase field, after suddenly switching
on the tunnel-coupling between a pair of quasi-1D condensates. These
'Quasibreathers' grow out of initial quantum fluctuations, mimicking
processes that are crucially important in other areas like cosmology.
They can be well described by exact analytical solutions of the sine-Gordon
model. We derived their mean lifetime and density after the system
reaches a statistical steady state. These predicitions are consistent
with our numerical simulations. An experimental realization, even
with present setups seems to be within reach. 

\emph{Acknowledgements}. \textendash{} We thank J. Schmiedmayer for
fruitful discussions. Financial support by the Emmy-Noether program
and the SFB/TR 12 is gratefully acknowledged. CN gratefully acknowledges
the hospitality of Boston University, where parts of this work were
performed. 

\bibliographystyle{apsrev4-1}
\bibliography{BibSG}

\begin{thebibliography}{19}%
\makeatletter
\providecommand \@ifxundefined [1]{%
 \@ifx{#1\undefined}
}%
\providecommand \@ifnum [1]{%
 \ifnum #1\expandafter \@firstoftwo
 \else \expandafter \@secondoftwo
 \fi
}%
\providecommand \@ifx [1]{%
 \ifx #1\expandafter \@firstoftwo
 \else \expandafter \@secondoftwo
 \fi
}%
\providecommand \natexlab [1]{#1}%
\providecommand \enquote  [1]{``#1''}%
\providecommand \bibnamefont  [1]{#1}%
\providecommand \bibfnamefont [1]{#1}%
\providecommand \citenamefont [1]{#1}%
\providecommand \href@noop [0]{\@secondoftwo}%
\providecommand \href [0]{\begingroup \@sanitize@url \@href}%
\providecommand \@href[1]{\@@startlink{#1}\@@href}%
\providecommand \@@href[1]{\endgroup#1\@@endlink}%
\providecommand \@sanitize@url [0]{\catcode `\\12\catcode `\$12\catcode
  `\&12\catcode `\#12\catcode `\^12\catcode `\_12\catcode `\%12\relax}%
\providecommand \@@startlink[1]{}%
\providecommand \@@endlink[0]{}%
\providecommand \url  [0]{\begingroup\@sanitize@url \@url }%
\providecommand \@url [1]{\endgroup\@href {#1}{\urlprefix }}%
\providecommand \urlprefix  [0]{URL }%
\providecommand \Eprint [0]{\href }%
\providecommand \doibase [0]{http://dx.doi.org/}%
\providecommand \selectlanguage [0]{\@gobble}%
\providecommand \bibinfo  [0]{\@secondoftwo}%
\providecommand \bibfield  [0]{\@secondoftwo}%
\providecommand \translation [1]{[#1]}%
\providecommand \BibitemOpen [0]{}%
\providecommand \bibitemStop [0]{}%
\providecommand \bibitemNoStop [0]{.\EOS\space}%
\providecommand \EOS [0]{\spacefactor3000\relax}%
\providecommand \BibitemShut  [1]{\csname bibitem#1\endcsname}%
\let\auto@bib@innerbib\@empty
\bibitem [{\citenamefont {Gleiser}\ and\ \citenamefont
  {Howell}(2003)}]{2003_Gleiser}%
  \BibitemOpen
  \bibfield  {author} {\bibinfo {author} {\bibfnamefont {M.}~\bibnamefont
  {Gleiser}}\ and\ \bibinfo {author} {\bibfnamefont {R.~C.}\ \bibnamefont
  {Howell}},\ }\href@noop {} {\bibfield  {journal} {\bibinfo  {journal} {Phys.
  Rev. E}\ }\textbf {\bibinfo {volume} {68}},\ \bibinfo {pages} {065203}
  (\bibinfo {year} {2003})}\BibitemShut {NoStop}%
\bibitem [{\citenamefont {Amin}\ \emph {et~al.}(2011)\citenamefont {Amin},
  \citenamefont {Easther}, \citenamefont {Finkel}, \citenamefont {Flauger},\
  and\ \citenamefont {Hertzberg}}]{2011_AminInflation}%
  \BibitemOpen
  \bibfield  {author} {\bibinfo {author} {\bibfnamefont {M.~A.}\ \bibnamefont
  {Amin}}, \bibinfo {author} {\bibfnamefont {R.}~\bibnamefont {Easther}},
  \bibinfo {author} {\bibfnamefont {H.}~\bibnamefont {Finkel}}, \bibinfo
  {author} {\bibfnamefont {R.}~\bibnamefont {Flauger}}, \ and\ \bibinfo
  {author} {\bibfnamefont {M.~P.}\ \bibnamefont {Hertzberg}},\ }\href@noop {}
  {\bibfield  {journal} {\bibinfo  {journal} {arXiv:1106.3335v1}\ } (\bibinfo
  {year} {2011})}\BibitemShut {NoStop}%
\bibitem [{\citenamefont {Amin}\ \emph {et~al.}(2010)\citenamefont {Amin},
  \citenamefont {Easther},\ and\ \citenamefont {Finkel}}]{2010_Amin3D}%
  \BibitemOpen
  \bibfield  {author} {\bibinfo {author} {\bibfnamefont {M.~A.}\ \bibnamefont
  {Amin}}, \bibinfo {author} {\bibfnamefont {R.}~\bibnamefont {Easther}}, \
  and\ \bibinfo {author} {\bibfnamefont {H.}~\bibnamefont {Finkel}},\
  }\href@noop {} {\bibfield  {journal} {\bibinfo  {journal} {Journal of
  Cosmology and Astroparticle Physics}\ }\textbf {\bibinfo {volume} {12}}
  (\bibinfo {year} {2010})}\BibitemShut {NoStop}%
\bibitem [{\citenamefont {Farhi}\ \emph {et~al.}(2008)\citenamefont {Farhi},
  \citenamefont {Graham}, \citenamefont {Guth}, \citenamefont {Iqbal},
  \citenamefont {Rosales},\ and\ \citenamefont {Stamatopoulos}}]{2008_Farhi}%
  \BibitemOpen
  \bibfield  {author} {\bibinfo {author} {\bibfnamefont {E.}~\bibnamefont
  {Farhi}}, \bibinfo {author} {\bibfnamefont {N.}~\bibnamefont {Graham}},
  \bibinfo {author} {\bibfnamefont {A.~H.}\ \bibnamefont {Guth}}, \bibinfo
  {author} {\bibfnamefont {N.}~\bibnamefont {Iqbal}}, \bibinfo {author}
  {\bibfnamefont {R.~R.}\ \bibnamefont {Rosales}}, \ and\ \bibinfo {author}
  {\bibfnamefont {N.}~\bibnamefont {Stamatopoulos}},\ }\href@noop {} {\bibfield
   {journal} {\bibinfo  {journal} {Phys. Rev. D}\ }\textbf {\bibinfo {volume}
  {77}},\ \bibinfo {pages} {085019} (\bibinfo {year} {2008})}\BibitemShut
  {NoStop}%
\bibitem [{\citenamefont {Hofferberth}\ \emph {et~al.}(2007)\citenamefont
  {Hofferberth}, \citenamefont {Lesanovsky}, \citenamefont {Fischer},
  \citenamefont {Schumm},\ and\ \citenamefont
  {Schmiedmayer}}]{2007_Hofferbeth}%
  \BibitemOpen
  \bibfield  {author} {\bibinfo {author} {\bibfnamefont {S.}~\bibnamefont
  {Hofferberth}}, \bibinfo {author} {\bibfnamefont {I.}~\bibnamefont
  {Lesanovsky}}, \bibinfo {author} {\bibfnamefont {B.}~\bibnamefont {Fischer}},
  \bibinfo {author} {\bibfnamefont {T.}~\bibnamefont {Schumm}}, \ and\ \bibinfo
  {author} {\bibfnamefont {J.}~\bibnamefont {Schmiedmayer}},\ }\href@noop {}
  {\bibfield  {journal} {\bibinfo  {journal} {Nature}\ }\textbf {\bibinfo
  {volume} {449}},\ \bibinfo {pages} {324} (\bibinfo {year}
  {2007})}\BibitemShut {NoStop}%
\bibitem [{\citenamefont {Betz}\ \emph {et~al.}(2011)\citenamefont {Betz},
  \citenamefont {Manz}, \citenamefont {B\"ucker}, \citenamefont {Berrada},
  \citenamefont {Koller}, \citenamefont {Kazakov}, \citenamefont {Mazets},
  \citenamefont {Stimming}, \citenamefont {Perrin}, \citenamefont {Schumm},\
  and\ \citenamefont {Schmiedmayer}}]{2011_Betz}%
  \BibitemOpen
  \bibfield  {author} {\bibinfo {author} {\bibfnamefont {T.}~\bibnamefont
  {Betz}}, \bibinfo {author} {\bibfnamefont {S.}~\bibnamefont {Manz}}, \bibinfo
  {author} {\bibfnamefont {R.}~\bibnamefont {B\"ucker}}, \bibinfo {author}
  {\bibfnamefont {T.}~\bibnamefont {Berrada}}, \bibinfo {author} {\bibfnamefont
  {C.}~\bibnamefont {Koller}}, \bibinfo {author} {\bibfnamefont
  {G.}~\bibnamefont {Kazakov}}, \bibinfo {author} {\bibfnamefont {I.~E.}\
  \bibnamefont {Mazets}}, \bibinfo {author} {\bibfnamefont {H.-P.}\
  \bibnamefont {Stimming}}, \bibinfo {author} {\bibfnamefont {A.}~\bibnamefont
  {Perrin}}, \bibinfo {author} {\bibfnamefont {T.}~\bibnamefont {Schumm}}, \
  and\ \bibinfo {author} {\bibfnamefont {J.}~\bibnamefont {Schmiedmayer}},\
  }\href@noop {} {\bibfield  {journal} {\bibinfo  {journal} {Phys. Rev. Lett.}\
  }\textbf {\bibinfo {volume} {106}},\ \bibinfo {pages} {020407} (\bibinfo
  {year} {2011})}\BibitemShut {NoStop}%
\bibitem [{\citenamefont {Gritsev}\ \emph
  {et~al.}(2007{\natexlab{a}})\citenamefont {Gritsev}, \citenamefont
  {Polkovnikov},\ and\ \citenamefont {Demler}}]{2007_AnatoliSG}%
  \BibitemOpen
  \bibfield  {author} {\bibinfo {author} {\bibfnamefont {V.}~\bibnamefont
  {Gritsev}}, \bibinfo {author} {\bibfnamefont {A.}~\bibnamefont
  {Polkovnikov}}, \ and\ \bibinfo {author} {\bibfnamefont {E.}~\bibnamefont
  {Demler}},\ }\href {\doibase 10.1103/PhysRevB.75.174511} {\bibfield
  {journal} {\bibinfo  {journal} {Phys. Rev. B}\ }\textbf {\bibinfo {volume}
  {75}},\ \bibinfo {pages} {174511} (\bibinfo {year}
  {2007}{\natexlab{a}})}\BibitemShut {NoStop}%
\bibitem [{\citenamefont {Gritsev}\ \emph
  {et~al.}(2007{\natexlab{b}})\citenamefont {Gritsev}, \citenamefont {Demler},
  \citenamefont {Lukin},\ and\ \citenamefont {Polkovnikov}}]{2007_GritsevPRL}%
  \BibitemOpen
  \bibfield  {author} {\bibinfo {author} {\bibfnamefont {V.}~\bibnamefont
  {Gritsev}}, \bibinfo {author} {\bibfnamefont {E.}~\bibnamefont {Demler}},
  \bibinfo {author} {\bibfnamefont {M.}~\bibnamefont {Lukin}}, \ and\ \bibinfo
  {author} {\bibfnamefont {A.}~\bibnamefont {Polkovnikov}},\ }\href@noop {}
  {\bibfield  {journal} {\bibinfo  {journal} {Phys. Rev. Lett.}\ }\textbf
  {\bibinfo {volume} {99}},\ \bibinfo {pages} {200404} (\bibinfo {year}
  {2007}{\natexlab{b}})}\BibitemShut {NoStop}%
\bibitem [{\citenamefont {De~Grandi}\ \emph {et~al.}(2008)\citenamefont
  {De~Grandi}, \citenamefont {Barankov},\ and\ \citenamefont
  {Polkovnikov}}]{2008_Grandi}%
  \BibitemOpen
  \bibfield  {author} {\bibinfo {author} {\bibfnamefont {C.}~\bibnamefont
  {De~Grandi}}, \bibinfo {author} {\bibfnamefont {R.~A.}\ \bibnamefont
  {Barankov}}, \ and\ \bibinfo {author} {\bibfnamefont {A.}~\bibnamefont
  {Polkovnikov}},\ }\href@noop {} {\bibfield  {journal} {\bibinfo  {journal}
  {Phys. Rev. Lett.}\ }\textbf {\bibinfo {volume} {101}},\ \bibinfo {pages}
  {230402} (\bibinfo {year} {2008})}\BibitemShut {NoStop}%
\bibitem [{\citenamefont {Lancaster}\ \emph {et~al.}(2010)\citenamefont
  {Lancaster}, \citenamefont {Gull},\ and\ \citenamefont
  {Mitra}}]{2010_MitraSG}%
  \BibitemOpen
  \bibfield  {author} {\bibinfo {author} {\bibfnamefont {J.}~\bibnamefont
  {Lancaster}}, \bibinfo {author} {\bibfnamefont {E.}~\bibnamefont {Gull}}, \
  and\ \bibinfo {author} {\bibfnamefont {A.}~\bibnamefont {Mitra}},\
  }\href@noop {} {\bibfield  {journal} {\bibinfo  {journal} {Phys. Rev. B}\
  }\textbf {\bibinfo {volume} {82}},\ \bibinfo {pages} {235124} (\bibinfo
  {year} {2010})}\BibitemShut {NoStop}%
\bibitem [{\citenamefont {Iucci}\ and\ \citenamefont
  {Cazalilla}(2010)}]{2010_CazalillaNJP}%
  \BibitemOpen
  \bibfield  {author} {\bibinfo {author} {\bibfnamefont {A.}~\bibnamefont
  {Iucci}}\ and\ \bibinfo {author} {\bibfnamefont {M.~A.}\ \bibnamefont
  {Cazalilla}},\ }\href@noop {} {\bibfield  {journal} {\bibinfo  {journal} {New
  Journal of Physics}\ }\textbf {\bibinfo {volume} {12}},\ \bibinfo {pages}
  {055019} (\bibinfo {year} {2010})}\BibitemShut {NoStop}%
\bibitem [{\citenamefont {Foster}\ \emph {et~al.}(2010)\citenamefont {Foster},
  \citenamefont {Yuzbashyan},\ and\ \citenamefont {Altshuler}}]{2010_Foster}%
  \BibitemOpen
  \bibfield  {author} {\bibinfo {author} {\bibfnamefont {M.~S.}\ \bibnamefont
  {Foster}}, \bibinfo {author} {\bibfnamefont {E.~A.}\ \bibnamefont
  {Yuzbashyan}}, \ and\ \bibinfo {author} {\bibfnamefont {B.~L.}\ \bibnamefont
  {Altshuler}},\ }\href@noop {} {\bibfield  {journal} {\bibinfo  {journal}
  {Phys. Rev. Lett.}\ }\textbf {\bibinfo {volume} {105}},\ \bibinfo {pages}
  {135701} (\bibinfo {year} {2010})}\BibitemShut {NoStop}%
\bibitem [{\citenamefont {Greene}\ \emph {et~al.}(1999)\citenamefont {Greene},
  \citenamefont {Kofman},\ and\ \citenamefont {Starobinsky}}]{1999_Greene}%
  \BibitemOpen
  \bibfield  {author} {\bibinfo {author} {\bibfnamefont {P.~B.}\ \bibnamefont
  {Greene}}, \bibinfo {author} {\bibfnamefont {L.}~\bibnamefont {Kofman}}, \
  and\ \bibinfo {author} {\bibfnamefont {A.~A.}\ \bibnamefont {Starobinsky}},\
  }\href@noop {} {\bibfield  {journal} {\bibinfo  {journal} {Nuclear Physics
  B}\ ,\ \bibinfo {pages} {423}} (\bibinfo {year} {1999})}\BibitemShut
  {NoStop}%
\bibitem [{\citenamefont {Bouchoule}(2005)}]{2005_Bochoule}%
  \BibitemOpen
  \bibfield  {author} {\bibinfo {author} {\bibfnamefont {I.}~\bibnamefont
  {Bouchoule}},\ }\href@noop {} {\bibfield  {journal} {\bibinfo  {journal}
  {Eur. Phys. J. D}\ ,\ \bibinfo {pages} {147}} (\bibinfo {year}
  {2005})}\BibitemShut {NoStop}%
\bibitem [{\citenamefont {McLaughlin}\ and\ \citenamefont
  {Scott}(1978)}]{1978_Scott}%
  \BibitemOpen
  \bibfield  {author} {\bibinfo {author} {\bibfnamefont {D.~W.}\ \bibnamefont
  {McLaughlin}}\ and\ \bibinfo {author} {\bibfnamefont {A.~C.}\ \bibnamefont
  {Scott}},\ }\href@noop {} {\bibfield  {journal} {\bibinfo  {journal} {Phys.
  Rev. A}\ }\textbf {\bibinfo {volume} {18}},\ \bibinfo {pages} {1652}
  (\bibinfo {year} {1978})}\BibitemShut {NoStop}%
\bibitem [{\citenamefont {Polkovnikov}(2010)}]{2010_AnatoliTWA}%
  \BibitemOpen
  \bibfield  {author} {\bibinfo {author} {\bibfnamefont {A.}~\bibnamefont
  {Polkovnikov}},\ }\href@noop {} {\bibfield  {journal} {\bibinfo  {journal}
  {Annals of Phys.}\ }\textbf {\bibinfo {volume} {325}} (\bibinfo {year}
  {2010})}\BibitemShut {NoStop}%
\bibitem [{\citenamefont {Whitlock}\ and\ \citenamefont
  {Bouchoule}(2003)}]{2003_Bochoule}%
  \BibitemOpen
  \bibfield  {author} {\bibinfo {author} {\bibfnamefont {N.~K.}\ \bibnamefont
  {Whitlock}}\ and\ \bibinfo {author} {\bibfnamefont {I.}~\bibnamefont
  {Bouchoule}},\ }\href {\doibase 10.1103/PhysRevA.68.053609} {\bibfield
  {journal} {\bibinfo  {journal} {Phys. Rev. A}\ }\textbf {\bibinfo {volume}
  {68}},\ \bibinfo {pages} {053609} (\bibinfo {year} {2003})}\BibitemShut
  {NoStop}%
\bibitem [{\citenamefont {Kr\"uger}\ \emph {et~al.}(2010)\citenamefont
  {Kr\"uger}, \citenamefont {Hofferberth}, \citenamefont {Mazets},
  \citenamefont {Lesanovsky},\ and\ \citenamefont
  {Schmiedmayer}}]{2010_KruegerPRL}%
  \BibitemOpen
  \bibfield  {author} {\bibinfo {author} {\bibfnamefont {P.}~\bibnamefont
  {Kr\"uger}}, \bibinfo {author} {\bibfnamefont {S.}~\bibnamefont
  {Hofferberth}}, \bibinfo {author} {\bibfnamefont {I.~E.}\ \bibnamefont
  {Mazets}}, \bibinfo {author} {\bibfnamefont {I.}~\bibnamefont {Lesanovsky}},
  \ and\ \bibinfo {author} {\bibfnamefont {J.}~\bibnamefont {Schmiedmayer}},\
  }\href@noop {} {\bibfield  {journal} {\bibinfo  {journal} {Phys. Rev. Lett.}\
  }\textbf {\bibinfo {volume} {105}},\ \bibinfo {pages} {265302} (\bibinfo
  {year} {2010})}\BibitemShut {NoStop}%
\bibitem [{\citenamefont {Maki}\ and\ \citenamefont
  {Takayama}(1979)}]{1979_Maki2}%
  \BibitemOpen
  \bibfield  {author} {\bibinfo {author} {\bibfnamefont {K.}~\bibnamefont
  {Maki}}\ and\ \bibinfo {author} {\bibfnamefont {H.}~\bibnamefont
  {Takayama}},\ }\href@noop {} {\bibfield  {journal} {\bibinfo  {journal}
  {Phys. Rev. B}\ }\textbf {\bibinfo {volume} {20}},\ \bibinfo {pages} {5002}
  (\bibinfo {year} {1979})}\BibitemShut {NoStop}%
\end{thebibliography}%

\clearpage{}

\appendix

\subsection*{Supplementary Material}

\emph{Derivation of Quasibreather Solution} \textendash{} Here, we
sketch the crucial steps involved in the derivation of the quasibreather
solutions discussed in the main text. The details of the rather lengthy
calculation will be shown elsewhere. 

In general, the Bäcklund transform allows for the construction of
arbitrary multisoliton waves for the classical sine-Gordon equation
(SGE) $\phi_{tt}-\phi_{xx}+\sin\phi=0$ (setting for the moment $\beta=m=1$).
The idea is that starting from an arbitrary solution $\psi$ of the
SGE, one can easily show that the field $\phi$ fulfilling the Bäcklund
equations (BE's) 
\begin{eqnarray}
\frac{1}{2}\frac{\partial}{\partial z}(\phi-\psi) & = & \alpha\sin\left[\frac{\phi+\psi}{2}\right],\label{Eq1}
\end{eqnarray}
 
\begin{eqnarray}
\frac{1}{2}\frac{\partial}{\partial\tau}(\phi+\psi) & = & \frac{1}{\alpha}\sin\left[\frac{\phi-\psi}{2}\right],\label{Eq2}
\end{eqnarray}
where $\alpha\in\mathbb{C}$ is an arbitary constant, $z\equiv\frac{x-t}{2}$
and $\tau\equiv\frac{x+t}{2}$, is a solution as well. Furthermore,
it turns out that $\phi$ obtained in this manner is just the original
solution plus an additional soliton (see for instance \cite{1978_Scott}).
For real values of $\alpha$, the velocity of the added soliton is
given by $v=\frac{\alpha^{2}-1}{\alpha^{2}+1}$. In the considered
case, the known solution $\psi$ is the homogeneous part of the phase-field,
$\Phi(t)$, performing Josephson oscillations according to $\ddot{\Phi}=-\sin\Phi$.
As the numerical analysis with the TWA shows, slight inhomogeneties
in the field are amplified parametrically, eventually leading to the
formation of (standing) breather-like excitations, which have a finite
lifetime. We construct exact solutions of the SGE describing standing
breathers on top of this oscillating 'background'. As it was shown
in the main publication, it turns out that the solutions constructed
here are well suited to describe the observed localized patterns in
the field $\phi$.

In the absence of the background, ordinary breathers of the SGE can
be obtained by adding a soliton and an antisoliton to the 'vaccum'.
Here, we follow closely this strategy. For the purpose of creating
a single soliton, the parameter $\alpha$ can be choosen to be \emph{real}.
However, in order to create finally a breather out of a soliton-antisoliton
pair, we must allow for $\alpha\in\mathbb{C}$. Being interested in
standing quasibreathers (breather + background), it turns out that
in a first step, formally, one has to add a single 'soliton' with
$|\alpha|=1$ on top of the background. Solving Eqs.$\,$(\ref{Eq1},\ref{Eq2})
with $\psi\equiv\Phi$, one obtains a set of solutions $\phi_{\alpha}(x,t)$.
Fortunately, for the second step (adding the corresponding anti-soliton
$\phi_{\beta}$), one can use the fact \cite{1978_Scott} that 
\begin{eqnarray}
\chi(x,t) & = & 4\arctan\left[\frac{\alpha+\beta}{\alpha-\beta}\tan\left(\frac{\phi_{\alpha}-\phi_{\beta}}{4}\right)\right]+\Phi(t),\label{Eq3-1}
\end{eqnarray}
 is a solution of the SGE as well, as long as both $\phi_{\alpha}$
and $\phi_{\beta}$ were obtained from Eqs.$\,$(\ref{Eq1},\ref{Eq2})
with $\psi\equiv\Phi$. The first step (finding the solution for a
'soliton' on top of an oscillating background) involves solving the
set of coupled differential equations Eqs.$\,$(\ref{Eq1},\ref{Eq2}).
The second step mainly consists in matching $\alpha$ and $\beta$
properly (it turns out that $\alpha=\beta^{\ast}$). 

Finally, one finds that the quasibreather solutions $\chi$ have the
form 
\begin{eqnarray}
\chi & = & \frac{4}{\beta}\arctan\left(\frac{2}{\tan\varphi}\frac{{\rm Im}\mathcal{F}[x,t;\Phi_{0},\Delta_{0},\varphi]}{1+|\mathcal{F}|^{2}}\right)+\Phi.\label{eq0-1}
\end{eqnarray}
The argument of the $\arctan$-function corresponds to the function
$\mathcal{G}$ introduced in the main text. In fact, there are \emph{two}
sets of solutions: 
\begin{eqnarray}
\frac{\mathcal{F}}{f} & = & \left\{ \begin{array}{ccc}
\frac{\tanh[\frac{mfx}{4}+\mathcal{C}+i\Delta]+\mathcal{A}}{\mathcal{B}} & , & \cos\varphi>\sin\left(\frac{\beta\Phi_{0}}{2}\right)\\
\frac{-\tan[\frac{mfx}{4}+\mathcal{C}+i\Delta]+\mathcal{A}}{\mathcal{B}} & , & \cos\varphi\leq\sin\left(\frac{\beta\Phi_{0}}{2}\right)
\end{array}\right.\label{Solution2}
\end{eqnarray}
Whereas the well known, standing SGM-breather is characterized by
a single parameter (its amplitude), the solution here depends on three
parameters. The amplitude of the underlying background oscillations
$\Phi_{0}$, the maximal amplitude of the breather (determined by
a parameter $\varphi\in[0,\pi/2]$), and the relative phase between
breather and background oscillations, set by $\Delta_{0}$. 

The function $\Delta(t)$ in Eq.~(\ref{Solution2}) grows monotonically
during the temporal evolution of $\chi$. It is given by the integral
equation 
\begin{eqnarray}
\Delta(t) & = & \frac{mf\sin(2\varphi)}{2}\int_{0}^{t}dt'\,\frac{\sin^{2}[\beta\Phi(t')/2]}{|\mathcal{B}|^{2}(t')}+\Delta_{0}.\label{eq:1}
\end{eqnarray}
Furthermore, $\mathcal{B}(t)\equiv\beta m^{-1}\dot{\Phi}+2i\sin\varphi\sin(\beta\Phi/2)$
and $\mathcal{A}(t)\equiv2f^{-1}\cos\varphi\cos(\beta\Phi(t)/2)$,
where $f\equiv2|\cos^{2}\varphi-\sin^{2}(\beta\Phi_{0}/2)|^{1/2}$.
The solutions for $\cos\varphi>\sin(\beta\Phi_{0}/2)$ are periodic
in time {[}with $\mathcal{C}(t)=\tanh^{-1}(\sqrt{\mathcal{A}^{2}-1}-\mathcal{A})${]}.
These are termed 'single quasibreathers' and denoted by $\chi_{\varphi}$.
For $\cos\varphi\leq\sin(\beta\Phi_{0}/2)$, $\chi$ is periodic in
space. These 'multibreather' solutions $\chi_{k}$ are labeled by
the wavenumber $k=mf/2$ {[}in this case $\mathcal{C}(t)=\tan^{-1}(\sqrt{\mathcal{A}^{2}+1}+\mathcal{A})${]}.

\begin{figure}
\includegraphics[width=1\columnwidth]{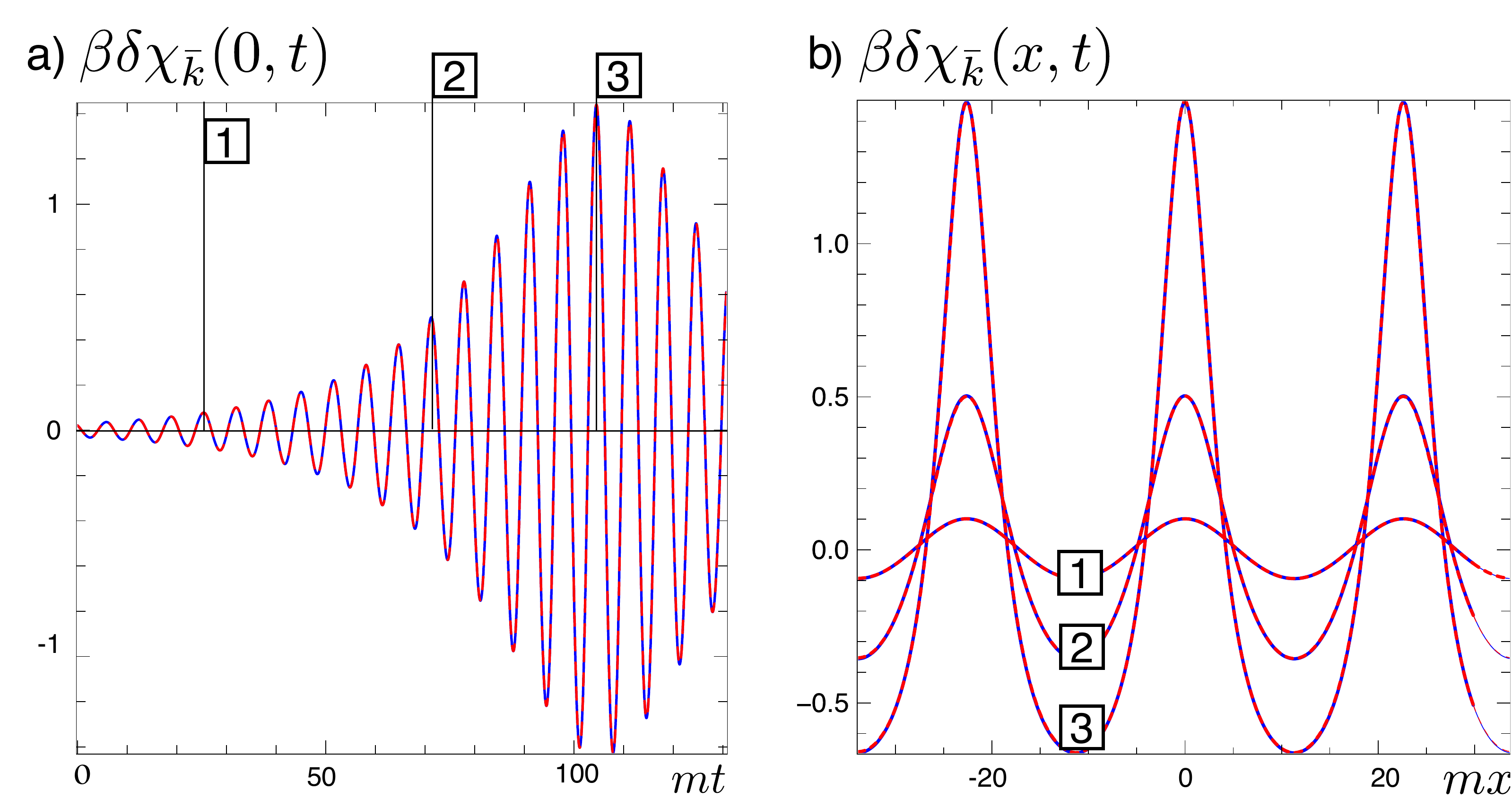} \caption{a) 'Multibreather' solution $\chi_{k}$ with $k=m\beta\Phi_{0}/\sqrt{8}$
and $\beta\Phi_{0}=0.25\pi$, evaluated at the center of the excitation
at $x=0$. We plot $\delta\chi_{k}=\chi_{k}-L^{-1}\int_{0}^{L}\, dx\chi_{k}$
($L$ denotes the system size). The inhomogenity $\delta\phi$ is
amplified parametrically. The blue, solid lines show the analytical
result. b) Cut in $x$-direction for the times indicated in a). One
observes the transition from initial, sinusoidal inhomogenities into
the characteristic breather-peaks at large times. The analytical solution
$\chi_{k}$ is checked against a direct simulation of the sine-Gordon
equation with a 4th-order Runge-Kutta algorithm (with appropriate
initial conditions; red, dashed line) \label{Fig1}}
\end{figure}

From now on, we consider $\beta\Phi_{0}\lesssim1$. Averaging $d\Delta/dt$
over one Josephson period, one obtains $2\Delta\simeq\delta\Omega t+2\Delta_{0}$,
with a frequency difference (see main text) 
\begin{eqnarray}
\delta\Omega(\varphi,\Phi_{0}) & = & \frac{mf\cos\varphi}{2(1+\sin\varphi)}.\label{eq:2}
\end{eqnarray}
The 'multibreather' solution $\chi_{k}$ has remarkable properties
(see Fig.$\,$\ref{Fig1}). For $t\rightarrow-\infty$ {[}$\Delta\rightarrow-\infty${]}
it reduces to a phonon $\delta\chi_{k}$ with wavenumber $k$, superimposed
on the oscillating background. It is amplified parametrically, yielding
\begin{eqnarray}
\delta\chi_{k} & \propto & e^{\Gamma_{k}t}{\rm \zeta}(t)\cos kx,\label{Eq3-1}
\end{eqnarray}
where $\zeta(t)$ is a periodic function. The amplification rate $\Gamma_{k}=\delta\Omega(\varphi(k),\Phi_{0})$.
As soon as $\Delta\simeq-1$, the characteristic breather peaks form
and exist during a time set by $\tau_{k}=4/\Gamma_{k}$. The maximum
is reached for $\Delta\simeq0$, where one observes a set of QBs with
distance $2\pi/k$. For $\Delta\rightarrow\infty$, these decay again.
In constrast, $\chi_{\varphi}$ describes a single QB whose amplitude
is modulated quasiperiodically with period $2\pi/\delta\Omega(\varphi,\Phi_{0})$. 

\emph{Correlation Function}. \textendash{}At long times $t/\bar{\tau}\gg1$,
as reported in the main text, one observes a steady state characterized
by the perpetual creation and annihilation of breathers. In order
to confirm the presence of quasibreathers in the experiment, we propose
to evaluate the phase-phase correlation function $C_{\phi\phi}(x)\equiv\langle\hat{\phi}(x,t)\hat{\phi}(0,t)\rangle-\langle\hat{\phi}(x,t)\rangle\langle\hat{\phi}(0,t)\rangle$.
Note that as we restrict to initial detunings $\beta\Phi_{0}\lesssim1$,
we do not observe the creation of solitons after the quench and the
corresponding phase slips in $\phi$. Therefore, we do not have to
switch to correlation functions like $\langle\cos\beta\hat{\phi}(x,t)\cos\beta\hat{\phi}(x,0)\rangle$. 

One can find a decent analytical approximation for $C_{\phi\phi}$
based on the 'multibreather' solutions discussed above. The 'multibreather'
solution is strictly periodic in space. However, note that this exact
solution of the classical sine-Gordon is only valid when the breather
amplitude is large enough compared to the noisy background. We assume
that each of these quasibreathers centered at $x_{i}$ (their phases
are encoded in the time $t_{i}$) is described by
\begin{eqnarray}
\tilde{\chi}(x;x_{i},t_{i}) & \equiv & \Theta(\frac{\bar{\lambda}}{2}-|x-x_{i}|)\Theta(\frac{\bar{\tau}}{2}-|t_{i}|)\chi_{\bar{k}}(x-x_{i},t_{i}).\label{eq:-2}
\end{eqnarray}
Here, $\chi_{\bar{k}}$ denotes the 'multibreather' solution with
$k=\bar{k}$ and $\Theta$ is the heavy-side step function. For simplicity,
we pick the special set of solutions with $\Delta_{0}=0$ (at small
values of $\beta\Phi_{0}$, $\chi_{\bar{k}}$ depends only slightly
on the precise relative phase between breather and background, $\Delta_{0}$,
anyway). With this choice of $\Delta_{0}$, the 'multibreather' solution
$\chi_{\bar{k}}(x,t)$ reaches its maximal value at $t=0$ and $x=n\bar{\lambda}$
with $n\in\mathbb{Z}$. After some time, a 'real' quasibreather dives
into the noisy background and the 'multibreather' solution ceases
to be valid. While it is hard to determine the precise time when this
happens, in a reasonable approximation, we cut-off $\tilde{\chi}(x;x_{i},t_{i})$
for $|t_{i}|>\bar{\tau}/2$. Here, $\bar{\tau}$ is the 'lifetime'
of the 'multibreather' solution (see main text). 

In a given experimental run $\alpha$, $\phi_{\alpha}(x,t)\simeq\sum_{j=1}^{n_{\alpha}}\tilde{\chi}(x;x_{i}^{\alpha},t_{i}^{\alpha})$
with $n_{\alpha}$ denoting the number of breathers at time $t$ (neglecting
short-wavelength radiation). It is understood that the distance between
neighboring breathers $|x_{i}-x_{i+1}|\simeq\bar{\lambda}$. Under
the assumption of independent quasibreathers and with $\langle\hat{\phi}\rangle\simeq0$
as well as $C_{\phi\phi}(x,t)\simeq C_{\phi\phi}(x)$ for $t/\bar{\tau}\gg1$,
one obtains

\begin{eqnarray}
C_{\phi\phi}(x) & \simeq & \overline{\sum_{i=1}^{n_{\alpha}}\tilde{\chi}(x;x_{i}^{\alpha},t_{i}^{\alpha})\tilde{\chi}(0;x_{i}^{\alpha},t_{i}^{\alpha})}.\label{eq:-1}
\end{eqnarray}
The bar denotes the average over experimental runs. Note that only
a single quasibreather with $|x_{i_{0}}^{\alpha}|\leq\bar{\lambda}/2$
contributes to the average and that the correlation function vanishes
in this approximation of independent breathers for $|x|>\bar{\lambda}$.
Therefore, we can replace the average over runs by an average over
$x_{0}\equiv x_{i_{0}}^{\alpha}$ and $t_{0}\equiv t_{i_{0}}^{\alpha}$
yielding
\begin{eqnarray}
 &  & C_{\phi\phi}(x)\nonumber \\
 & \simeq & \frac{1}{\bar{\tau}\bar{\lambda}}\int_{-\bar{\lambda}/2}^{\bar{\lambda}/2}dx_{0}\int_{-\bar{\tau}/2}^{\bar{\tau}/2}dt\,\tilde{\chi}(x;x_{0},t_{0})\tilde{\chi}(0;x_{0},t_{0}),\label{eq:-3}
\end{eqnarray}
which can be evaluated numerically.
\end{document}